\documentclass[reprint,prl]{revtex4-2}

\usepackage[colorlinks=true, allcolors=blue]{hyperref}
\usepackage{graphicx}% Include figure files
\usepackage{dcolumn}% Align table columns on decimal point
\usepackage{bm}% bold math
\usepackage{siunitx}
\usepackage{xcolor}
\usepackage{braket}
\usepackage{mathtools}
\usepackage{amsmath}
\usepackage{leftidx}
\usepackage{wasysym}
\usepackage{float}
\usepackage{cleveref}
\Crefname{equation}{Eq.}{Eqs.}

\usepackage{amsfonts} %% <- also included by amssymb
\DeclareMathSymbol{\shortminus}{\mathbin}{AMSa}{"39}
\usepackage{physics}

\usepackage{mypackage}
\usepackage{numericalvalues}
\usepackage{beamdivergence}

\usepackage[T1]{fontenc}
\DeclareSIUnit[number-unit-product = {}]{\inch}{\text{\textquotedbl}}

\graphicspath{{images/}}

\begin{document}

%TC:ignore

\title{Transverse beam emittance measurement by undulator radiation power noise}

\author{Ihar Lobach}
\email{ilobach@uchicago.edu}
\affiliation{The University of Chicago, Department of Physics, Chicago, IL
60637, USA}%Lines break automatically or can be forced with \\
 
\author{Sergei Nagaitsev}
\altaffiliation[Also at ]{The Enrico Fermi Institute, The University of Chicago,
Chicago, Illinois 60637, USA}
\author{Valeri Lebedev}
\author{Aleksandr Romanov}
\author{Giulio Stancari} 
\author{Alexander Valishev} 
\affiliation{Fermi National Accelerator Laboratory, Batavia, IL 60510, USA}%
\author{Aliaksei Halavanau}
\author{Zhirong Huang}
\affiliation{SLAC National Accelerator Laboratory, Stanford University, Menlo
Park CA 94025, USA}
\author{Kwang-Je Kim}
\altaffiliation[Also at ]{The Enrico Fermi Institute, The University of Chicago,
Chicago, Illinois 60637, USA} \affiliation{Argonne National Accelerator
Laboratory, Lemont, IL 60439, USA}%

\begin{abstract}
% must be < 600 characters
Generally, turn-to-turn power fluctuations of incoherent spontaneous synchrotron radiation in a
storage ring depend on the 6D phase-space distribution of the electron bunch. In
some cases, if only one parameter of the distribution is unknown, this parameter
can be determined from the measured magnitude of these power fluctuations. In
this Letter, we report an absolute measurement (no free parameters or
calibration) of a small vertical emittance (5--15 nm rms) of a flat beam by this
method, under conditions, when it is unresolvable by a conventional synchrotron
light beam size monitor.
\end{abstract}

\maketitle

%TC:endignore

\begin{figure*}[t]
    \includegraphics[width=\textwidth]{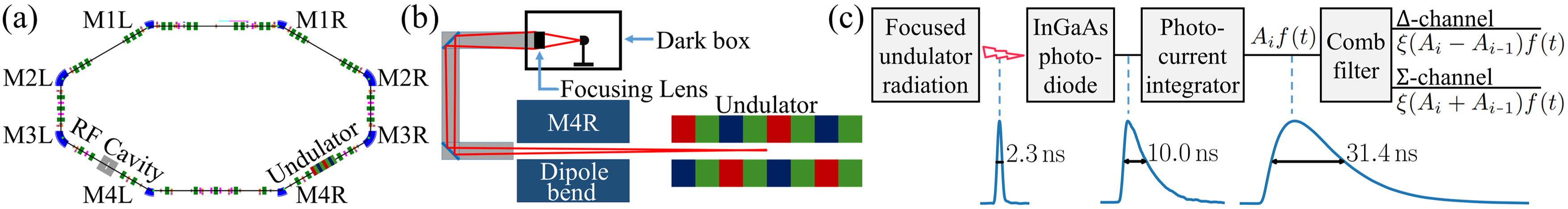}
    \caption{\label{fig:iota_detector_layout} (a) Layout of IOTA, electrons circulate clockwise. (b) Light path
    from the undulator to the detector  (not to scale). (c) Block diagram of the
    detection circuit, $i$ is the IOTA revolution number. The delay in the comb
    filter equals exactly one IOTA revolution, $\valrevolutionPeriod$. Hence, its
    \D-channel provides a signal difference between two consecutive turns.}
\end{figure*}

Most often, noise is encountered in a negative context and is considered
something that needs to be minimized. However, there are multiple examples where
noise is used as a non-invasive probe into the parameters of a certain system,
and even to measure fundamental constants. Examples include the determination of
the Boltzmann constant $k_B$ by the thermal noise in an electrical conductor
\cite{johnson1928thermal} and the measurement of the elementary charge $e$ by
the shot noise of the electric current in a vacuum tube
\cite{hull1925determination}. In fact, the latter effect is also relevant to
accelerators and storage rings, where it is known as Schottky noise
\cite{vanderMeer} due to the finite number of charge carriers in the beam, as
described  by  Schottky \cite{schottky}. Many beam parameters, such as
the momentum spread, the number of particles and even transverse rms emittances,
are imprinted into the power spectrum of Schottky noise. It is often used in
beam diagnostics \cite{boussard1986schottky,van1989diagnostics,caspers20074}.

Synchrotron radiation is generated by individual electrons in the beam.
Hence, Schottky noise in the beam current must pass on to the synchrotron
radiation power in some way. Therefore, one could assume that the synchrotron
radiation power noise may carry information about beam parameters as well. This assumption is, in fact, correct. Three decades ago,
Ref.~\cite{teich1990statistical} reported the results of an experimental study
into statistical properties of wiggler radiation in a storage ring. It was noted
that the magnitude of turn-to-turn intensity fluctuations depends on the
dimensions of the electron bunch. The potential in beam instrumentation was soon
realized \cite{zolotorev1996fluctuational} and a number of papers followed.
However, to this day, mostly measurements of a bunch length via these
fluctuations were discussed \cite{sannibale2009absolute,sajaev2004measurement,
sajaev2000determination}. Only Ref.~\cite{catravas1999measurement} reported an
order-of-magnitude measurement of a transverse emittance. In this Letter, we
describe a new fluctuations-based technique for an absolute measurement of a
transverse emittance. There are no free parameters in our equations, nor is a
calibration required. However, the transverse and longitudinal focusing functions of the storage ring are assumed to be known. This technique is tested at the Integrable Optics Test Accelerator (IOTA) storage ring at Fermilab \cite{antipov2017iota}. For a beam
with approximately equal and relatively large transverse rms emittances, the results agree with conventional visible synchrotron light monitors (SLMs)
\cite{kuklev2019synchrotron}. Then, in a different regime, we measure a much
smaller vertical emittance of a flat beam, unresolvable by our SLMs. These emittance measurements agree with estimates, based on the beam lifetime. We also
discuss possible further improvements.

Let us assume that we have a detector that can measure the number of detected
synchrotron radiation photons $\N$ at each revolution in a storage ring. Then,
according to \cite{lobach2020PRAB,kim2017synchrotron,teich1990statistical,park2019investigation}, the
variance of this number is 
\begin{equation}\label{eq:varN_from_book}
    \var{\N}=\av{\big(\N-\av{\N}\bigr)^2}=\av{\N}+\frac{1}{M}\av{\N}^2,
\end{equation}
\noindent where the linear term represents the photon shot noise, related to the
quantum discrete nature of light. This effect would exist even if there was only
one electron, circulating in the ring. Indeed, the electron would radiate photons with a Poisson
distribution~\cite{glauber1963quantum,glauber1963coherent,glauber1951some}. The quadratic term in \Cref{eq:varN_from_book} corresponds to the interference of fields, radiated by different electrons.
Changes in relative electron positions and velocities, inside the
bunch, result in fluctuations of the radiation power and, consequently, of the number of detected
photons. In a storage ring, the effect arises from betatron and synchrotron
motion, from radiation induced diffusion, etc. The dependence of $\var{\N}$ on the 6D
phase-space distribution of the electron bunch is introduced through the
parameter $M$, which is conventionally called the number of coherent modes \cite{kim2017synchrotron,teich1990statistical,lobach2020PRAB}. In
addition to bunch parameters, $M$ depends on the specific spectral-angular distribution of the
radiation, on the angular aperture, and on the detection efficiency (as a function of
wavelength). Previously, we derived an equation for $M$ \Meqref~for a Gaussian
transverse beam profile and an arbitrary longitudinal bunch density distribution
$\rho(z)$ (normalized), assuming an rms bunch length much longer than the
radiation wavelength. In this Letter, $M$ is calculated by this equation
numerically, using our computer code \cite{furrepo}, as a function of
transverse emittances $\ex$ and $\ey$, the rms momentum spread $\sp$, and the effective bunch length,
$\szeff=1/(2\sqrt{\pi}\int\rho^2(z)\dd{z})$, equal to the rms bunch length, $\sz$,
for a Gaussian distribution.

\newcommand{\tx}{\theta_x}
\newcommand{\ty}{\theta_y}
\newcommand{\stx}{\sigma_{\tx}}
\newcommand{\sty}{\sigma_{\ty}}

For illustration purposes, let us assume a Gaussian spectral-angular distribution for the number of detected photons $\N$, namely,
\begin{equation}\label{eq:gaussNdist}
    \frac{\dd[3]{\N}}{\dd k\dd\theta_x \dd\theta_y} = C \exp\bigl[
        -\frac{(k-k_0)^2}{2\usk^2}-\frac{\tx^2}{2\stx^2}-\frac{\ty^2}{2\sty^2}
        \bigr],
\end{equation}
\noindent where $k$ is the magnitude of the wave vector, $\tx$ and $\ty$ represent
the horizontal and vertical angles of the direction of the radiation in the
paraxial approximation, $k_0$ refers to the center of the radiation spectrum, $\usk$ is the spectral rms width, $\stx$ and $\sty$ are the angular rms
radiation sizes, $C$ is a constant. Then \cite{sannibale2009absolute,lobach2020furjointprab}
\begin{equation}\label{eq:MforwardIncoh}
    M = \sqrt{1+4\usk^2\sz^2}\sqrt{1+4k_0^2\sigma_{\theta_x}^2\sx^2}
    \sqrt{1+4k_0^2\sigma_{\theta_y}^2\sy^2},
\end{equation}
\noindent where $\sx$, $\sy$, $\sz$ are the rms sizes (determined by beam
emittances) of a Gaussian electron bunch. In addition, it is assumed that the radiation is
longitudinally incoherent $k_0\sz\gg1$ and that the radiation bandwidth is very
narrow $\usk\ll 1/(\sx\stx)$, $\usk\ll 1/(\sy\sty)$. In general, the
distribution parameters $k_0$, $\usk$, $\stx$, $\sty$ are determined by both the
properties of the emitted synchrotron radiation and by the properties of the
detecting system (angular aperture, detection efficiency). In \Cref{eq:MforwardIncoh}, the beam divergence is neglected and $M$
depends on $\sx$ and $\sy$, as opposed to a more general result \Meqref, where it depends on $\ex$ and $\ey$.

In our experiment, a single electron bunch circulated in IOTA with a revolution
period of $\valrevolutionPeriod$ and beam energy of $\valEbeamDimLess \pm
\valEbeamerror$.
We studied undulator radiation because the quadratic term in
\Cref{eq:varN_from_book}, sensitive to the bunch parameters, is larger for
undulators and wigglers than it is for dipole magnets \cite{lobach2020PRAB}.
The undulator parameter is $\Ku=\valKu$ with the number of periods
$\Nu=\valNundPer$ and the period length $\lamu=\SI{5.5}{cm}$.
The measurements were performed in the vicinity of the fundamental harmonic,
$\lambda_1=\lamu(1+\Ku^2/2)/(2\gamma^2)=\vallambdaone$, where $\gamma$ is the Lorentz factor.
As a photodetector we used an InGaAs PIN photodiode \cite{hamamatsu_photodiode}, whose quantum efficiency is about \SI{80}{\percent} around $\lambda_1$.
The photodiode was
installed in a dark box above the M4R dipole magnet, see Figs.~\ref{fig:iota_detector_layout}(a),(b). The light produced in the undulator exited the vacuum chamber through a window at the M4R dipole magnet. It was
directed to the box by a system of two mirrors, and focused by a lens
(focal distance $F = \SI{150}{mm}$)
into a spot, smaller than the photodiode's sensitive area ($\diameter\SI{1.0}{mm}$).
The lens was $\valzobs$ away from the center of the undulator.
We numerically calculated the spectral-angular distribution of the undulator 
radiation by our computer code \cite{wigradrepo}, based on the equations from 
\cite{clarke2004science}. 
Further, we used the manufacturers' specifications to account for the spectral
properties of the optical elements and the photodiode.
The resulting spectral width of the radiation was $\valfwhm$ (FWHM), and the  radiation was confined to a cone with a $\SI{2}{mrad}$ half angle. It could be fully transmitted through the $\diameter \SI{2}{\inch}$ optical system ($\SI{1}{\inch}/\valzobs=\valsemiApertureX$).
The simulated average number of detected photons per pass, per one electron of the electron bunch was
$\valphotonFluxTheorDimLess$. The empirical value was $\valphotonFluxMeasDimLess$ \cite{lobach2020furjointprab}. There were no
free adjustable parameters in this calculation.

Figure~\ref{fig:iota_detector_layout}(c) illustrates our
photodetection circuit. First, the radiation pulse is converted into a
photocurrent pulse by the photodiode. Then, the photocurrent pulse is integrated
by an op-amp-based RC integrator and converted to a voltage signal $A_i f(t)$, where $A_i$ is the signal amplitude at the $i$th turn and $f(t)$ is the average signal for one turn, normalized so that its maximum value is 1. The number of detected photons (photoelectrons) at the $i$th turn can be obtained as
\begin{equation}\label{eq:AtoN}
    \N_i = \chi A_i,
\end{equation}
\noindent where $\chi = \valAmpToPhotoel$, with a $\SI{5}{\percent}$ uncertainty, as per the characteristics of our
integrator and  the photodiode \cite{lobach2020furjointprab}. The
op-amp was capable of driving the 50-$\Omega$ input load of a fast digitizing scope,
located $\approx \SI{100}{m}$ away. In our measurements, $A_i\in[0,1.2]\SI{}{V}$.

The expected relative fluctuation of $A_i$
was \SIrange{e-4}{e-3}{} (rms), which is considerably lower than the
digitization resolution of our 8-bit broad-band oscilloscope. To overcome this
problem, we employed a passive comb (notch) filter \cite{smith2010physical}, see
Fig.~\ref{fig:iota_detector_layout}(c).
In this filter, the input signal first passes through a two-way splitter.
One arm is delayed relative to the other by exactly one IOTA revolution. Then, the difference and the sum
of the two signals are produced in the output \D- and \S-channels.
For an ideal comb filter,
\begin{align}
    \Dit{} = \xi(A_i-A_{i\shortminus 1})\ft{},\label{eq:Deq}\\
    \Sit{} = \xi(A_i+A_{i\shortminus 1})\ft{}\label{eq:Seq}.
\end{align}
In our filter, $\xi=\valxi$, which was measured by comparing input and output pulses. Now, since the offset was removed (\Cref{eq:Deq}), we were able to directly observe the sub-\SI{}{mV} turn-to-turn fluctuations in the \D-channel and the oscilloscope operated in the appropriate scale setting with negligible
digitization noise.

For each measurement, we recorded $\valwfLen$-long waveforms (about
$\valNrev$ IOTA revolutions) of \D- and \S-channels with the oscilloscope at
$\SI{20}{GSa/s}$. The beam current decay was negligible during the acquisition time. The
photoelectron count variance $\var{\N}$ and the photoelectron count mean
$\av{\N}$ were obtained from the $\valNrev$ collected amplitudes, $\Delta_i(\tpeak)$ and
$\Sigma_i(\tpeak)$, as
\begin{align}
    \var{\N} = \chi^2\var{A} = \frac{\chi^2 \var{\Delta(\tpeak)}}{2\xi^2},\label{eq:varNfromDamp}\\
    \av{\N} = \chi \av{A} = \chi\frac{\av{\Sigma(\tpeak)}}{2\xi}\label{eq:avNfromSamp},
\end{align}
\noindent  where $\tpeak$ is the time within each turn corresponding to the peak of the signal, $f(\tpeak)=1$. These formulas follow from \Cref{eq:AtoN,eq:Deq,eq:Seq}. There was a small cross-talk
($<\SI{1}{\percent}$) between the output channels of the comb filter. However, its effect is
negligible in \Cref{eq:varNfromDamp,eq:avNfromSamp}. Also, there was some
instrumental noise contribution to $\var{\Delta(\tpeak)}$. Its contribution to $\var{\N}$ was
$\valnoiseLevel$, as measured at zero beam current. Primary sources of
this noise were the integrator's op-amp and the oscilloscope's pre-amp. In
\cite{lobach2020furjointprab}, we showed that this noise level was independent of
$\av{\N}$ via measurements with an independent test light source. Therefore, it
can be simply subtracted. Reference~\cite{lobach2020furjointprab}
also describes the details of the photocurrent integrator and the comb filter.

The number of coherent modes $M$ and, hence, the
fluctuations $\var{\N}$ depend on the following bunch parameters: $\ex$,
$\ey$ (or mode emittances $\e_1$, $\e_2$), $\sp$, $\szeff$. When only one of them is unknown and
$\var{\N}$ is known (or measured), we can numerically solve \Cref{eq:varN_from_book}, using
our general formula for $M$ \Meqref, to find the unknown bunch parameter. Below we
consider two such situations.

\begin{figure*}[t]
    \includegraphics[width=\textwidth]{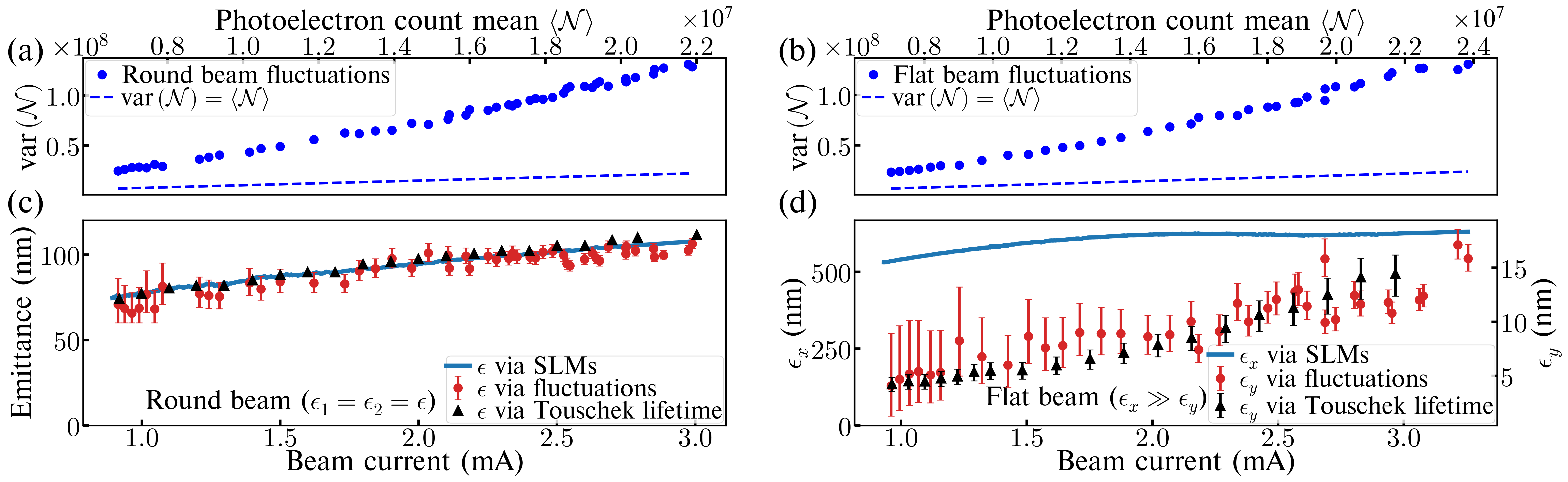}
    \caption{\label{fig:iota_measurements} Panels (a) and (b) show the measured
    fluctuations for the round and flat beams, respectively. The statistical error of each point is $\valvarNerror$ (not shown).  (c) The round-beam mode emittance $\e$, determined via SLMs, via undulator radiation
    fluctuations, and via Touschek lifetime, assuming the effective momentum acceptance $\valMomApEff$.
    (d) The flat-beam horizontal emittance measurement via SLMs (left
    scale), the vertical emittance measurement via fluctuations and via Touschek lifetime (right scale).
    The SLMs had a monitor-to-monitor spread of $\pm\valSLMroundErr$ (round beam) and $\pm\valSLMflatErr$ (horizontal emittance of flat beam); these error bars are not shown.
    All emittances are rms, unnormalized.}
\end{figure*}

In the first case, we consider a strongly coupled \cite{lebedev2010betatron,lebedev2020ibs} transverse
focusing optics in IOTA, which was specifically designed to keep the two mode emittances equal $\e_1=\e_2=\e$.
This was empirically confirmed to be true with a few percent precision.
We will call this setup ``round beam''.
The longitudinal bunch profile was
measured by a high-bandwidth wall-current monitor \cite{WGM} to determine
$\szeff$ and estimate $\sp$. The fluctuations $\var{\N}$, measured using \Cref{eq:varNfromDamp}, are shown in
Fig.~\ref{fig:iota_measurements}(a), with a statistical error of $\valvarNerror$ (at all beam currents), which was determined with an independent test light source \cite{lobach2020furjointprab}.
Hence, the only unknown parameter in \Cref{eq:varN_from_book} is $\e$.
The numerical solution of \Cref{eq:varN_from_book} with $M$ from
\Meqref~was performed on the Midway2 cluster at the University of Chicago
Research Computing Center. The results for $\e$ are shown in
Fig.~\ref{fig:iota_measurements}(c) (red points). The error bars correspond to the statistical error of the fluctuations measurement.
Apart from this statistical error there is also a systematic error due to the $\valEbeamerror$ uncertainty on the beam energy (from $\SI{10}{nm}$ at lower beam currents to $\SI{14}{nm}$ at higher currents).  

In IOTA, transverse beam sizes are monitored by seven SLMs, at
M1L--M4L and at M1R--M3R, see Fig.~\ref{fig:iota_detector_layout}(a). Beam
emittances can be determined from the measured sizes using the design Twiss functions. Such measurements for $\e$ of the round beam [blue line
in Fig.~\ref{fig:iota_measurements}(c)] agree with the fluctuations-based $\e$
within the uncertainties.
The smallest reliably resolvable emittance by the SLMs in our experiment configuration was $\approx\SI{20}{nm}$.
The measured round-beam emittance $\e$ is \SIrange{75}{100}{nm} (rms, unnormalized), primarily due to intrabeam scattering \cite{bjorken1982intrabeam, IBS_Nagaitsev}. The expected zero-current value is $\e \approx \SI{12}{nm}$.

In the second case, we consider uncoupled focusing, with the vertical emittance much smaller than the horizontal one. We will call this configuration ``flat beam''. The horizontal emittance $\ex$ of the flat beam can still be reliably measured by
the SLMs; $\szeff$ and $\sp$ can still be measured by the wall-current monitor.
However, the seven SLMs provided very inconsistent estimates for the much
smaller $\ey$ --- the max-to-min variation for different SLMs reached a factor of eight.
We believe this happened because the beam images were close to the resolution limit, set by a combination of factors, such as the diffraction limit, the point spread function of the cameras, chromatic aberrations, the effective radiator size of the dipole magnet radiation ($\approx\SI{20}{\micro m}$), and the camera pixel size ($\approx\SI{10}{\micro m}$ in terms of beam size).
Therefore, the monitor-to-monitor emittance variation primarily came from the Twiss beta-function variation ($\beta_y^{(\mathrm{max})}/\beta_y^{(\mathrm{min})} \approx 12$).
Although the resolution of the SLMs may be improved in the future \cite{lobach2020furjointprab}, at present, $\ey$ of the flat beam is unresolvable by the SLMs, and, therefore, is truly unknown. However, the measured fluctuations for the flat beam, shown in Fig.~\ref{fig:iota_measurements}(b), were of the same order as for the round beam, with the same statistical error. Hence, we were able to reconstruct $\ey$ in the same way as
$\e$ in Fig.~\ref{fig:iota_measurements}(c). The results are shown in
Fig.~\ref{fig:iota_measurements}(d) (red points, right vertical scale) along with the SLMs data for
$\ex$ (blue line, left vertical scale). In addition to the statistical error of $\ey$, shown in Fig.~\ref{fig:iota_measurements}(d), there was also a systematic error due to the $\valEbeamerror$ uncertainty on the beam energy (from $\SI{2.5}{nm}$ at lower currents to $\SI{5}{nm}$ at higher currents),
and a systematic error due to the $\valSLMflatErr$ uncertainty on $\ex$ (from $\SI{1.3}{nm}$ at lower currents to $\SI{2.4}{nm}$ at higher currents).
% These systematic errors are relatively high. However, they can be reduced by future improvements of beam characterization in IOTA.
The measured vertical emittance is \SIrange{5}{15}{nm}, most
likely due to a nonzero residual transverse coupling. The expected zero-current flat-beam emittances were $\ex \approx
\SI{50}{nm}$, $\ey \gtrsim \SI{0.33}{p m}$ (set by the quantum excitation in a perfectly uncoupled ring).

The vertical emittance $\ey$ of the flat beam in IOTA could also be estimated from the
measured beam lifetime $\abs{I/(\dv*{I}{t})}$, assuming that it is determined solely
by Touschek scattering \cite{touschek}, which is a good approximation at beam currents
$I\gtrsim\SI{0.5}{mA}$ \cite{lebedev2020ibs}. In storage rings, the Touschek lifetime is determined by the effective
momentum acceptance $\momApEff$ \cite{Carmignani:2014qyb}, which is smaller than or equal to the rf bucket half-height, $\momApRF=\valdppSep$ in IOTA. We measured the IOTA beam lifetime (\SIrange{550}{1000}{s}) for both round and flat beams as a function of beam current.  Using the known round-beam emittance and the bunch length, we arrived at the following estimate for IOTA, $\momApEff=\valMomApEff$,
by comparing the calculated \cite{Intro_touschek,piwinski1999touschek} Touschek lifetime and the measured beam lifetime 
(for details see Appendix~D of \cite{lobach2020furjointprab}).
The black triangles in Fig.~\ref{fig:iota_measurements}(c) illustrate the emittance of the round beam $\e$, determined from the measured beam lifetime using the Touschek lifetime calculation with $\momApEff=\valMomApEff$.
Then, we used this value of $\momApEff$ and the values of
$\ex$, measured by the SLMs, to estimate the vertical emittance $\ey$ of the flat beam via the Touschek lifetime. The results are shown in 
Fig.~\ref{fig:iota_measurements}(d) (black triangles). The error bars correspond to the $\pm\valSLMflatErr$ uncertainty on $\ex$.
% The agreement with the fluctuations-based measurement is rather good. 

During our measurements, the rms and the effective bunch lengths $\sz$, $\szeff$ were \SIrange{26}{31}{cm} and \SIrange{24}{30}{cm}, respectively, primarily due to intrabeam scattering. They were different because the longitudinal bunch shape was not exactly Gaussian due to beam interaction with its environment \cite{haissinski1973exact}.
The relative rms momentum spread was
$\sp \approx \valszTospDimLess\times\sz[\mathrm{cm}]$, based on the rf cavity and ring parameters. The expected zero-current
values are $\sz=\szeff=\valszZeroCur$, $\sp=\valspZeroCur$. The uncoupled case Twiss beta-functions in the undulator were $\bx=\valbetax$,
$\by=\valbetay$, for more details see \cite{lobach2020furjointprab}.

% the text below could potentially be deleted to make the paper shorter
Other emittance monitors (wire scanners, Compton-scattering
monitors \cite{tenenbaum1999measurement,sakai2001measurement}) could provide
better resolution in IOTA.
However, if a bright synchrotron light source is available, our fluctuations-based monitor
may be a good inexpensive non-invasive alternative.
% the text above could potentially be deleted to make the paper shorter
There are two requirements for the technique to work: (A) the fluctuations should not be dominated by the Poisson noise, so that $M$ can be reliably deduced
from $\var{\N}$, and (B) $M$ has to be sensitive to $\ex$, $\ey$.
Let us consider the $h$th harmonic of undulator radiation in the approximation of \Cref{eq:gaussNdist,eq:MforwardIncoh} with a narrow Gaussian filter $\usk\ll k_0/(h\Nu)$ and $k_0=2\pi h/\lambda_1$. By integrating
\Cref{eq:gaussNdist} we obtain $\av{\N}=C(2\pi)^{3/2}\stx\sty\usk$, where $C$ is the peak on-axis
photon flux, $C=\alpha\Nu^2\gamma^2F_h(\Ku)n_e/k_0$
\cite[p.~68]{clarke2004science}, $\alpha$ is the fine-structure constant,
$n_e$ is the number of electrons per bunch, and the function $F_h(\Ku)$, defined
in \cite[p.~69]{clarke2004science},
is typically about \SIrange{0.2}{0.4}{}. If we approximate \Cref{eq:MforwardIncoh} by $M \approx 8\, k_0^2 \, \usk\stx \sty \sx \sy \sz $, the requirement~(A) becomes [see \Cref{eq:varN_from_book}]
\begin{equation}\label{eq:avNMreq}
    \frac{\av{\N}}{M}=\alpha\left(\frac{\pi}{2}\right)^{\frac{3}{2}}
    F_h(\Ku)\frac{\gamma^2\Nu^2n_e}{\sx\sy\sz k_0^3} \gtrsim 1.
\end{equation}

In the model of \Cref{eq:MforwardIncoh}, the requirement~(B) becomes $\sx\gtrsim1/(2k_0\stx)$, $\sy\gtrsim1/(2k_0\sty)$.
Notably, one can intentionally make $M$ insensitive to $\sx$ (or $\sy$), and, thus, enable an independent measurement of $\sy$ (or $\sx$). For example, by using a vertical slit, which can be approximated by a very small $\stx\ll 1/(2k_0 \sx)$, one can deduce $\sy$ from a measured $M$ without the knowledge of $\sx$.
Also, radiation masks can be applied to analyze fluctuations in various portions of the angular distribution of the radiation, which adds flexibility to this method.
Assuming no angular restrictions, $\stx,\sty\approx\sqrt{\lambda_0/(2\Lu)}$ \cite[Eq.~(2.57)]{kim2017synchrotron}, and the requirement (B) becomes
\begin{equation}\label{eq:resolutionlimit}
    \sx,\sy \gtrsim \sqrt{2\Lu \lambda_0}/(4\pi),
\end{equation}
\noindent where $\lambda_0=2\pi/k_0$, $\Lu$ is the undulator length.
In IOTA, this corresponds to $\sx,\sy\gtrsim\valslim$, or $\ex\gtrsim\valexlim$,
$\ey\gtrsim\valeylim$; and $\av{\N}/M\in [2.3,5.0]$ (as per measurements).

Equation~\eqref{eq:resolutionlimit} shows that the resolution limit improves
with a shorter wavelength. Therefore, this technique may be particularly
beneficial for existing state-of-the-art and next generation low-emittance high-brightness
ultraviolet and x-ray synchrotron light sources. Consider the Advanced Photon
Source Upgrade (APS-U) with a round beam configuration for example. The beam energy is \SI{6}{GeV}, $n_e=\SI{9.6e10}{}$, $\sz=\SI{3.1}{cm}$, $\ex=\SI{31.9}{pm}$,
$\ey=\SI{31.7}{pm}$, $\sx=\SI{12.9}{\micro m}$,
$\sy=\SI{8.7}{\micro m}$, $\sxp=\SI{2.5}{\micro rad}$, $\syp=\SI{3.6}{\micro
rad}$ \cite{apsu}. Let us use the fundamental harmonic $\lambda_1=\SI{4.1}{\angstrom}$ of the undulator with $\lamu=\SI{28}{mm}$, $\Ku=\SI{2.459}{}$, and $\Lu=\SI{2.1}{m}$. Equation~\eqref{eq:avNMreq} yields $\av{\N}/M=\SI{19}{}$, and \Cref{eq:resolutionlimit} becomes $\sx,\sy\gtrsim\SI{3.3}{\micro m}$.
Thus, both requirements~(A) and (B) are satisfied.
% All assumptions leading to \Cref{eq:resolutionlimit,eq:avNMreq} are fulfilled. 
These estimates were confirmed by \cite[Eqs.~(2--8)]{lobach2020furjointprab}.
\begin{acknowledgments}
We would like to thank the entire FAST/IOTA team at Fermilab for helping us with
building and installing the experimental setup and taking data, especially Mark
Obrycki, James Santucci, and Wayne Johnson. Greg Saewert constructed the detection circuit and provided the test light source.
Brian Fellenz, Daniil Frolov, David Johnson, and Todd Johnson provided equipment and assisted during our
detector tests.
This work was completed in part with resources provided by the University of Chicago Research Computing Center. This research is supported by the University of Chicago and the US Department of Energy under contracts DE-AC02-76SF00515 and
DE-AC02-06CH11357. This manuscript has been authored by Fermi Research Alliance,
LLC under Contract No. DE-AC02-07CH11359 with the U.S. Department of Energy,
Office of Science, Office of High Energy Physics.
\end{acknowledgments}
%TC:endignore

\bibliography{bibliography}

\begin{thebibliography}{40}%
\makeatletter
\providecommand \@ifxundefined [1]{%
 \@ifx{#1\undefined}
}%
\providecommand \@ifnum [1]{%
 \ifnum #1\expandafter \@firstoftwo
 \else \expandafter \@secondoftwo
 \fi
}%
\providecommand \@ifx [1]{%
 \ifx #1\expandafter \@firstoftwo
 \else \expandafter \@secondoftwo
 \fi
}%
\providecommand \natexlab [1]{#1}%
\providecommand \enquote  [1]{``#1''}%
\providecommand \bibnamefont  [1]{#1}%
\providecommand \bibfnamefont [1]{#1}%
\providecommand \citenamefont [1]{#1}%
\providecommand \href@noop [0]{\@secondoftwo}%
\providecommand \href [0]{\begingroup \@sanitize@url \@href}%
\providecommand \@href[1]{\@@startlink{#1}\@@href}%
\providecommand \@@href[1]{\endgroup#1\@@endlink}%
\providecommand \@sanitize@url [0]{\catcode `\\12\catcode `\$12\catcode
  `\&12\catcode `\#12\catcode `\^12\catcode `\_12\catcode `\%12\relax}%
\providecommand \@@startlink[1]{}%
\providecommand \@@endlink[0]{}%
\providecommand \url  [0]{\begingroup\@sanitize@url \@url }%
\providecommand \@url [1]{\endgroup\@href {#1}{\urlprefix }}%
\providecommand \urlprefix  [0]{URL }%
\providecommand \Eprint [0]{\href }%
\providecommand \doibase [0]{https://doi.org/}%
\providecommand \selectlanguage [0]{\@gobble}%
\providecommand \bibinfo  [0]{\@secondoftwo}%
\providecommand \bibfield  [0]{\@secondoftwo}%
\providecommand \translation [1]{[#1]}%
\providecommand \BibitemOpen [0]{}%
\providecommand \bibitemStop [0]{}%
\providecommand \bibitemNoStop [0]{.\EOS\space}%
\providecommand \EOS [0]{\spacefactor3000\relax}%
\providecommand \BibitemShut  [1]{\csname bibitem#1\endcsname}%
\let\auto@bib@innerbib\@empty
%</preamble>
\bibitem [{\citenamefont {Johnson}(1928)}]{johnson1928thermal}%
  \BibitemOpen
  \bibfield  {author} {\bibinfo {author} {\bibfnamefont {J.~B.}\ \bibnamefont
  {Johnson}},\ }\bibfield  {title} {\bibinfo {title} {Thermal agitation of
  electricity in conductors},\ }\href@noop {} {\bibfield  {journal} {\bibinfo
  {journal} {Phys. Rev.}\ }\textbf {\bibinfo {volume} {32}},\ \bibinfo {pages}
  {97} (\bibinfo {year} {1928})}\BibitemShut {NoStop}%
\bibitem [{\citenamefont {Hull}\ and\ \citenamefont
  {Williams}(1925)}]{hull1925determination}%
  \BibitemOpen
  \bibfield  {author} {\bibinfo {author} {\bibfnamefont {A.~W.}\ \bibnamefont
  {Hull}}\ and\ \bibinfo {author} {\bibfnamefont {N.}~\bibnamefont
  {Williams}},\ }\bibfield  {title} {\bibinfo {title} {Determination of
  elementary charge {e} from measurements of shot-effect},\ }\href@noop {}
  {\bibfield  {journal} {\bibinfo  {journal} {Phys. Rev.}\ }\textbf {\bibinfo
  {volume} {25}},\ \bibinfo {pages} {147} (\bibinfo {year} {1925})}\BibitemShut
  {NoStop}%
\bibitem [{\citenamefont {van~der Meer}(1984)}]{vanderMeer}%
  \BibitemOpen
  \bibfield  {author} {\bibinfo {author} {\bibfnamefont {S.}~\bibnamefont
  {van~der Meer}},\ }\bibfield  {title} {\bibinfo {title} {Stochastic cooling
  and the accumulation of antiprotons},\ }\href
  {https://www.nobelprize.org/prizes/physics/1984/meer/lecture/} {\bibfield
  {journal} {\bibinfo  {journal} {Nobel Lecture}\ } (\bibinfo {year}
  {1984})}\BibitemShut {NoStop}%
\bibitem [{\citenamefont {Schottky}(1918)}]{schottky}%
  \BibitemOpen
  \bibfield  {author} {\bibinfo {author} {\bibfnamefont {W.}~\bibnamefont
  {Schottky}},\ }\bibfield  {title} {\bibinfo {title} {Über spontane
  {S}tromschwankungen in verschiedenen {E}lektrizitätsleitern},\ }\href
  {https://doi.org/10.1002/andp.19183622304} {\bibfield  {journal} {\bibinfo
  {journal} {Annalen der Physik}\ }\textbf {\bibinfo {volume} {362}},\ \bibinfo
  {pages} {541} (\bibinfo {year} {1918})}\BibitemShut {NoStop}%
\bibitem [{\citenamefont {Boussard}(1986)}]{boussard1986schottky}%
  \BibitemOpen
  \bibfield  {author} {\bibinfo {author} {\bibfnamefont {D.}~\bibnamefont
  {Boussard}},\ }\href {https://doi.org/10.5170/CERN-1987-003-V-2.416} {\emph
  {\bibinfo {title} {Schottky noise and beam transfer function diagnostics}}},\
  \bibinfo {type} {Tech. Rep.}\ \bibinfo {number} {CERN-SPS-86-11-ARF}\
  (\bibinfo {year} {1986})\BibitemShut {NoStop}%
\bibitem [{\citenamefont {van~der Meer}(1989)}]{van1989diagnostics}%
  \BibitemOpen
  \bibfield  {author} {\bibinfo {author} {\bibfnamefont {S.}~\bibnamefont
  {van~der Meer}},\ }\bibfield  {title} {\bibinfo {title} {Diagnostics with
  {Schottky} noise},\ }in\ \href@noop {} {\emph {\bibinfo {booktitle}
  {Frontiers of Particle Beams; Observation, Diagnosis and Correction}}}\
  (\bibinfo  {publisher} {Springer},\ \bibinfo {year} {1989})\ pp.\ \bibinfo
  {pages} {423--433}\BibitemShut {NoStop}%
\bibitem [{\citenamefont {Caspers}\ \emph {et~al.}(2007)\citenamefont
  {Caspers}, \citenamefont {Jimenez}, \citenamefont {Jones}, \citenamefont
  {Kroyer}, \citenamefont {Vuitton}, \citenamefont {Hamerla}, \citenamefont
  {Jansson}, \citenamefont {Misek}, \citenamefont {Pasquinelli}, \citenamefont
  {Seifrid} \emph {et~al.}}]{caspers20074}%
  \BibitemOpen
  \bibfield  {author} {\bibinfo {author} {\bibfnamefont {F.}~\bibnamefont
  {Caspers}}, \bibinfo {author} {\bibfnamefont {J.~M.}\ \bibnamefont
  {Jimenez}}, \bibinfo {author} {\bibfnamefont {O.~R.}\ \bibnamefont {Jones}},
  \bibinfo {author} {\bibfnamefont {T.}~\bibnamefont {Kroyer}}, \bibinfo
  {author} {\bibfnamefont {C.}~\bibnamefont {Vuitton}}, \bibinfo {author}
  {\bibfnamefont {T.~W.}\ \bibnamefont {Hamerla}}, \bibinfo {author}
  {\bibfnamefont {A.}~\bibnamefont {Jansson}}, \bibinfo {author} {\bibfnamefont
  {J.}~\bibnamefont {Misek}}, \bibinfo {author} {\bibfnamefont {R.~J.}\
  \bibnamefont {Pasquinelli}}, \bibinfo {author} {\bibfnamefont
  {P.}~\bibnamefont {Seifrid}}, \emph {et~al.},\ }\bibfield  {title} {\bibinfo
  {title} {The 4.8 {GHz LHC Schottky} pick-up system},\ }in\ \href@noop {}
  {\emph {\bibinfo {booktitle} {2007 IEEE Particle Accelerator Conference
  (PAC)}}}\ (\bibinfo {organization} {IEEE},\ \bibinfo {year} {2007})\ pp.\
  \bibinfo {pages} {4174--4176}\BibitemShut {NoStop}%
\bibitem [{\citenamefont {Teich}\ \emph {et~al.}(1990)\citenamefont {Teich},
  \citenamefont {Tanabe}, \citenamefont {Marshall},\ and\ \citenamefont
  {Galayda}}]{teich1990statistical}%
  \BibitemOpen
  \bibfield  {author} {\bibinfo {author} {\bibfnamefont {M.~C.}\ \bibnamefont
  {Teich}}, \bibinfo {author} {\bibfnamefont {T.}~\bibnamefont {Tanabe}},
  \bibinfo {author} {\bibfnamefont {T.~C.}\ \bibnamefont {Marshall}},\ and\
  \bibinfo {author} {\bibfnamefont {J.}~\bibnamefont {Galayda}},\ }\bibfield
  {title} {\bibinfo {title} {Statistical properties of wiggler and
  bending-magnet radiation from the {Brookhaven Vacuum-Ultraviolet} electron
  storage ring},\ }\href@noop {} {\bibfield  {journal} {\bibinfo  {journal}
  {Phys. Rev. Lett.}\ }\textbf {\bibinfo {volume} {65}},\ \bibinfo {pages}
  {3393} (\bibinfo {year} {1990})}\BibitemShut {NoStop}%
\bibitem [{\citenamefont {Zolotorev}\ and\ \citenamefont
  {Stupakov}(1996)}]{zolotorev1996fluctuational}%
  \BibitemOpen
  \bibfield  {author} {\bibinfo {author} {\bibfnamefont {M.~S.}\ \bibnamefont
  {Zolotorev}}\ and\ \bibinfo {author} {\bibfnamefont {G.~V.}\ \bibnamefont
  {Stupakov}},\ }\href {http://cds.cern.ch/record/305348} {\emph {\bibinfo
  {title} {{Fluctuational interferometry for measurement of short pulses of
  incoherent radiation}}}},\ \bibinfo {type} {Tech. Rep.}\ \bibinfo {number}
  {SLAC-PUB-7132}\ (\bibinfo  {institution} {SLAC},\ \bibinfo {address}
  {Stanford, CA},\ \bibinfo {year} {1996})\BibitemShut {NoStop}%
\bibitem [{\citenamefont {Sannibale}\ \emph {et~al.}(2009)\citenamefont
  {Sannibale}, \citenamefont {Stupakov}, \citenamefont {Zolotorev},
  \citenamefont {Filippetto},\ and\ \citenamefont
  {J{\"a}gerhofer}}]{sannibale2009absolute}%
  \BibitemOpen
  \bibfield  {author} {\bibinfo {author} {\bibfnamefont {F.}~\bibnamefont
  {Sannibale}}, \bibinfo {author} {\bibfnamefont {G.}~\bibnamefont {Stupakov}},
  \bibinfo {author} {\bibfnamefont {M.}~\bibnamefont {Zolotorev}}, \bibinfo
  {author} {\bibfnamefont {D.}~\bibnamefont {Filippetto}},\ and\ \bibinfo
  {author} {\bibfnamefont {L.}~\bibnamefont {J{\"a}gerhofer}},\ }\bibfield
  {title} {\bibinfo {title} {Absolute bunch length measurements by incoherent
  radiation fluctuation analysis},\ }\href@noop {} {\bibfield  {journal}
  {\bibinfo  {journal} {Phys. Rev. ST Accel. Beams}\ }\textbf {\bibinfo
  {volume} {12}},\ \bibinfo {pages} {032801} (\bibinfo {year}
  {2009})}\BibitemShut {NoStop}%
\bibitem [{\citenamefont {Sajaev}(2004)}]{sajaev2004measurement}%
  \BibitemOpen
  \bibfield  {author} {\bibinfo {author} {\bibfnamefont {V.}~\bibnamefont
  {Sajaev}},\ }\bibfield  {title} {\bibinfo {title} {Measurement of bunch
  length using spectral analysis of incoherent radiation fluctuations},\ }in\
  \href@noop {} {\emph {\bibinfo {booktitle} {AIP Conf. Proc.}}},\ Vol.\
  \bibinfo {volume} {732}\ (\bibinfo  {publisher} {AIP},\ \bibinfo {year}
  {2004})\ pp.\ \bibinfo {pages} {73--87}\BibitemShut {NoStop}%
\bibitem [{\citenamefont {Sajaev}(2000)}]{sajaev2000determination}%
  \BibitemOpen
  \bibfield  {author} {\bibinfo {author} {\bibfnamefont {V.}~\bibnamefont
  {Sajaev}},\ }\href@noop {} {\emph {\bibinfo {title} {Determination of
  longitudinal bunch profile using spectral fluctuations of incoherent
  radiation}}},\ \bibinfo {type} {Report No}\ \bibinfo {number}
  {ANL/ASD/CP-100935}\ (\bibinfo  {institution} {Argonne National Laboratory},\
  \bibinfo {year} {2000})\BibitemShut {NoStop}%
\bibitem [{\citenamefont {Catravas}\ \emph {et~al.}(1999)\citenamefont
  {Catravas}, \citenamefont {Leemans}, \citenamefont {Wurtele}, \citenamefont
  {Zolotorev}, \citenamefont {Babzien}, \citenamefont {Ben-Zvi}, \citenamefont
  {Segalov}, \citenamefont {Wang},\ and\ \citenamefont
  {Yakimenko}}]{catravas1999measurement}%
  \BibitemOpen
  \bibfield  {author} {\bibinfo {author} {\bibfnamefont {P.}~\bibnamefont
  {Catravas}}, \bibinfo {author} {\bibfnamefont {W.}~\bibnamefont {Leemans}},
  \bibinfo {author} {\bibfnamefont {J.}~\bibnamefont {Wurtele}}, \bibinfo
  {author} {\bibfnamefont {M.}~\bibnamefont {Zolotorev}}, \bibinfo {author}
  {\bibfnamefont {M.}~\bibnamefont {Babzien}}, \bibinfo {author} {\bibfnamefont
  {I.}~\bibnamefont {Ben-Zvi}}, \bibinfo {author} {\bibfnamefont
  {Z.}~\bibnamefont {Segalov}}, \bibinfo {author} {\bibfnamefont {X.-J.}\
  \bibnamefont {Wang}},\ and\ \bibinfo {author} {\bibfnamefont
  {V.}~\bibnamefont {Yakimenko}},\ }\bibfield  {title} {\bibinfo {title}
  {Measurement of electron-beam bunch length and emittance using
  shot-noise-driven fluctuations in incoherent radiation},\ }\href@noop {}
  {\bibfield  {journal} {\bibinfo  {journal} {Phys. Rev. Lett.}\ }\textbf
  {\bibinfo {volume} {82}},\ \bibinfo {pages} {5261} (\bibinfo {year}
  {1999})}\BibitemShut {NoStop}%
\bibitem [{\citenamefont {Antipov}\ \emph {et~al.}(2017)\citenamefont
  {Antipov}, \citenamefont {Broemmelsiek}, \citenamefont {Bruhwiler},
  \citenamefont {Edstrom}, \citenamefont {Harms}, \citenamefont {Lebedev},
  \citenamefont {Leibfritz}, \citenamefont {Nagaitsev}, \citenamefont {Park},
  \citenamefont {Piekarz} \emph {et~al.}}]{antipov2017iota}%
  \BibitemOpen
  \bibfield  {author} {\bibinfo {author} {\bibfnamefont {S.}~\bibnamefont
  {Antipov}}, \bibinfo {author} {\bibfnamefont {D.}~\bibnamefont
  {Broemmelsiek}}, \bibinfo {author} {\bibfnamefont {D.}~\bibnamefont
  {Bruhwiler}}, \bibinfo {author} {\bibfnamefont {D.}~\bibnamefont {Edstrom}},
  \bibinfo {author} {\bibfnamefont {E.}~\bibnamefont {Harms}}, \bibinfo
  {author} {\bibfnamefont {V.}~\bibnamefont {Lebedev}}, \bibinfo {author}
  {\bibfnamefont {J.}~\bibnamefont {Leibfritz}}, \bibinfo {author}
  {\bibfnamefont {S.}~\bibnamefont {Nagaitsev}}, \bibinfo {author}
  {\bibfnamefont {C.-S.}\ \bibnamefont {Park}}, \bibinfo {author}
  {\bibfnamefont {H.}~\bibnamefont {Piekarz}}, \emph {et~al.},\ }\bibfield
  {title} {\bibinfo {title} {{IOTA (Integrable Optics Test Accelerator):}
  facility and experimental beam physics program},\ }\href@noop {} {\bibfield
  {journal} {\bibinfo  {journal} {J. Instrum.}\ }\textbf {\bibinfo {volume}
  {12}}\bibinfo  {number} { (03)},\ \bibinfo {pages} {T03002}}\BibitemShut
  {NoStop}%
\bibitem [{\citenamefont {Kuklev}\ \emph {et~al.}(2019)\citenamefont {Kuklev},
  \citenamefont {Jarvis}, \citenamefont {Kim}, \citenamefont {Romanov},
  \citenamefont {Santucci},\ and\ \citenamefont
  {Stancari}}]{kuklev2019synchrotron}%
  \BibitemOpen
\bibfield  {number} {  }\bibfield  {author} {\bibinfo {author} {\bibfnamefont
  {N.}~\bibnamefont {Kuklev}}, \bibinfo {author} {\bibfnamefont
  {J.}~\bibnamefont {Jarvis}}, \bibinfo {author} {\bibfnamefont
  {Y.}~\bibnamefont {Kim}}, \bibinfo {author} {\bibfnamefont {A.}~\bibnamefont
  {Romanov}}, \bibinfo {author} {\bibfnamefont {J.}~\bibnamefont {Santucci}},\
  and\ \bibinfo {author} {\bibfnamefont {G.}~\bibnamefont {Stancari}},\
  }\bibfield  {title} {\bibinfo {title} {Synchrotron radiation beam diagnostics
  at {IOTA} --- commissioning performance and upgrade efforts},\ }in\ \href
  {https://doi.org/doi:10.18429/JACoW-IPAC2019-WEPGW103} {\emph {\bibinfo
  {booktitle} {Proc. 10th International Particle Accelerator Conference
  (IPAC'19), Melbourne, Australia, 19-24 May 2019}}},\ \bibinfo {series and
  number} {\bibinfo {series} {International Particle Accelerator Conference}\
  No.~\bibinfo {number} {10}}\ (\bibinfo  {publisher} {JACoW Publishing},\
  \bibinfo {address} {Geneva, Switzerland},\ \bibinfo {year} {2019})\ pp.\
  \bibinfo {pages} {2732--2735},\ \bibinfo {note}
  {https://doi.org/10.18429/JACoW-IPAC2019-WEPGW103}\BibitemShut {NoStop}%
\bibitem [{\citenamefont {Lobach}\ \emph
  {et~al.}(2020{\natexlab{a}})\citenamefont {Lobach}, \citenamefont {Lebedev},
  \citenamefont {Nagaitsev}, \citenamefont {Romanov}, \citenamefont {Stancari},
  \citenamefont {Valishev}, \citenamefont {Halavanau}, \citenamefont {Huang},\
  and\ \citenamefont {Kim}}]{lobach2020PRAB}%
  \BibitemOpen
  \bibfield  {author} {\bibinfo {author} {\bibfnamefont {I.}~\bibnamefont
  {Lobach}}, \bibinfo {author} {\bibfnamefont {V.}~\bibnamefont {Lebedev}},
  \bibinfo {author} {\bibfnamefont {S.}~\bibnamefont {Nagaitsev}}, \bibinfo
  {author} {\bibfnamefont {A.}~\bibnamefont {Romanov}}, \bibinfo {author}
  {\bibfnamefont {G.}~\bibnamefont {Stancari}}, \bibinfo {author}
  {\bibfnamefont {A.}~\bibnamefont {Valishev}}, \bibinfo {author}
  {\bibfnamefont {A.}~\bibnamefont {Halavanau}}, \bibinfo {author}
  {\bibfnamefont {Z.}~\bibnamefont {Huang}},\ and\ \bibinfo {author}
  {\bibfnamefont {K.-J.}\ \bibnamefont {Kim}},\ }\bibfield  {title} {\bibinfo
  {title} {Statistical properties of spontaneous synchrotron radiation with
  arbitrary degree of coherence},\ }\href
  {https://doi.org/10.1103/PhysRevAccelBeams.23.090703} {\bibfield  {journal}
  {\bibinfo  {journal} {Phys. Rev. Accel. Beams}\ }\textbf {\bibinfo {volume}
  {23}},\ \bibinfo {pages} {090703} (\bibinfo {year}
  {2020}{\natexlab{a}})}\BibitemShut {NoStop}%
\bibitem [{\citenamefont {Kim}\ \emph {et~al.}(2017)\citenamefont {Kim},
  \citenamefont {Huang},\ and\ \citenamefont {Lindberg}}]{kim2017synchrotron}%
  \BibitemOpen
  \bibfield  {author} {\bibinfo {author} {\bibfnamefont {K.-J.}\ \bibnamefont
  {Kim}}, \bibinfo {author} {\bibfnamefont {Z.}~\bibnamefont {Huang}},\ and\
  \bibinfo {author} {\bibfnamefont {R.}~\bibnamefont {Lindberg}},\ }\href@noop
  {} {\emph {\bibinfo {title} {Synchrotron radiation and free-electron
  lasers}}}\ (\bibinfo  {publisher} {Cambridge University Press},\ \bibinfo
  {year} {2017})\BibitemShut {NoStop}%
\bibitem [{\citenamefont {Park}(2019)}]{park2019investigation}%
  \BibitemOpen
  \bibfield  {author} {\bibinfo {author} {\bibfnamefont {J.-W.}\ \bibnamefont
  {Park}},\ }\emph {\bibinfo {title} {An Investigation of Possible Non-Standard
  Photon Statistics in a Free-Electron Laser}},\ \href
  {http://hdl.handle.net/10125/66233} {Ph.D. thesis},\ \bibinfo  {school}
  {University of Hawaii at Manoa} (\bibinfo {year} {2019})\BibitemShut
  {NoStop}%
\bibitem [{\citenamefont {Glauber}(1963{\natexlab{a}})}]{glauber1963quantum}%
  \BibitemOpen
  \bibfield  {author} {\bibinfo {author} {\bibfnamefont {R.~J.}\ \bibnamefont
  {Glauber}},\ }\bibfield  {title} {\bibinfo {title} {The quantum theory of
  optical coherence},\ }\href@noop {} {\bibfield  {journal} {\bibinfo
  {journal} {Phys. Rev.}\ }\textbf {\bibinfo {volume} {130}},\ \bibinfo {pages}
  {2529} (\bibinfo {year} {1963}{\natexlab{a}})}\BibitemShut {NoStop}%
\bibitem [{\citenamefont {Glauber}(1963{\natexlab{b}})}]{glauber1963coherent}%
  \BibitemOpen
  \bibfield  {author} {\bibinfo {author} {\bibfnamefont {R.~J.}\ \bibnamefont
  {Glauber}},\ }\bibfield  {title} {\bibinfo {title} {Coherent and incoherent
  states of the radiation field},\ }\href@noop {} {\bibfield  {journal}
  {\bibinfo  {journal} {Phys. Rev.}\ }\textbf {\bibinfo {volume} {131}},\
  \bibinfo {pages} {2766} (\bibinfo {year} {1963}{\natexlab{b}})}\BibitemShut
  {NoStop}%
\bibitem [{\citenamefont {Glauber}(1951)}]{glauber1951some}%
  \BibitemOpen
  \bibfield  {author} {\bibinfo {author} {\bibfnamefont {R.~J.}\ \bibnamefont
  {Glauber}},\ }\bibfield  {title} {\bibinfo {title} {Some notes on
  multiple-boson processes},\ }\href@noop {} {\bibfield  {journal} {\bibinfo
  {journal} {Phys. Rev.}\ }\textbf {\bibinfo {volume} {84}},\ \bibinfo {pages}
  {395} (\bibinfo {year} {1951})}\BibitemShut {NoStop}%
\bibitem [{\citenamefont {Lobach}\ \emph
  {et~al.}(2020{\natexlab{b}})\citenamefont {Lobach}, \citenamefont
  {Nagaitsev}, \citenamefont {Lebedev}, \citenamefont {Romanov}, \citenamefont
  {Stancari}, \citenamefont {Valishev}, \citenamefont {Halavanau},
  \citenamefont {Huang},\ and\ \citenamefont {Kim}}]{lobach2020furjointprab}%
  \BibitemOpen
  \bibfield  {author} {\bibinfo {author} {\bibfnamefont {I.}~\bibnamefont
  {Lobach}}, \bibinfo {author} {\bibfnamefont {S.}~\bibnamefont {Nagaitsev}},
  \bibinfo {author} {\bibfnamefont {V.}~\bibnamefont {Lebedev}}, \bibinfo
  {author} {\bibfnamefont {A.}~\bibnamefont {Romanov}}, \bibinfo {author}
  {\bibfnamefont {G.}~\bibnamefont {Stancari}}, \bibinfo {author}
  {\bibfnamefont {A.}~\bibnamefont {Valishev}}, \bibinfo {author}
  {\bibfnamefont {A.}~\bibnamefont {Halavanau}}, \bibinfo {author}
  {\bibfnamefont {Z.}~\bibnamefont {Huang}},\ and\ \bibinfo {author}
  {\bibfnamefont {K.-J.}\ \bibnamefont {Kim}},\ }\href@noop {} {\bibinfo
  {title} {Measurements of undulator radiation power noise and comparison with
  \textit{ab initio} calculations}} (\bibinfo {year} {2020}{\natexlab{b}}),\
  \Eprint {https://arxiv.org/abs/2012.00965} {arXiv:2012.00965
  [physics.acc-ph]} \BibitemShut {NoStop}%
\bibitem [{\citenamefont {Lobach}(2020{\natexlab{a}})}]{furrepo}%
  \BibitemOpen
  \bibfield  {author} {\bibinfo {author} {\bibfnamefont {I.}~\bibnamefont
  {Lobach}},\ }\href@noop {} {\bibinfo {title} {The source code for calculation
  of fluctuations in wiggler radiation}},\ \bibinfo {howpublished}
  {\url{https://github.com/IharLobach/fur}} (\bibinfo {year}
  {2020}{\natexlab{a}})\BibitemShut {NoStop}%
\bibitem [{ham()}]{hamamatsu_photodiode}%
  \BibitemOpen
  \href@noop {} {\bibinfo {title} {{Hamamatsu InGaAs PIN photodiode
  G11193-10R}}},\ \bibinfo {howpublished}
  {\url{https://www.hamamatsu.com/us/en/product/type/G11193-10R/index.html}},\
  \bibinfo {note} {accessed: 2020-11-18}\BibitemShut {NoStop}%
\bibitem [{\citenamefont {Lobach}(2020{\natexlab{b}})}]{wigradrepo}%
  \BibitemOpen
  \bibfield  {author} {\bibinfo {author} {\bibfnamefont {I.}~\bibnamefont
  {Lobach}},\ }\href@noop {} {\bibinfo {title} {The source code for calculation
  of spectral-angular distribution of wiggler radiation}},\ \bibinfo
  {howpublished} {\url{https://github.com/IharLobach/wigrad}} (\bibinfo {year}
  {2020}{\natexlab{b}})\BibitemShut {NoStop}%
\bibitem [{\citenamefont {Clarke}(2004)}]{clarke2004science}%
  \BibitemOpen
  \bibfield  {author} {\bibinfo {author} {\bibfnamefont {J.~A.}\ \bibnamefont
  {Clarke}},\ }\bibinfo {title} {The science and technology of undulators and
  wigglers}\ (\bibinfo  {publisher} {Oxford University Press on Demand},\
  \bibinfo {year} {2004})\ pp.\ \bibinfo {pages} {66--67}\BibitemShut {NoStop}%
\bibitem [{\citenamefont {Smith}(2010)}]{smith2010physical}%
  \BibitemOpen
  \bibfield  {author} {\bibinfo {author} {\bibfnamefont {J.~O.}\ \bibnamefont
  {Smith}},\ }\href@noop {} {\emph {\bibinfo {title} {Physical audio signal
  processing: For virtual musical instruments and audio effects}}}\ (\bibinfo
  {publisher} {W3K publishing},\ \bibinfo {year} {2010})\BibitemShut {NoStop}%
\bibitem [{\citenamefont {Lebedev}\ and\ \citenamefont
  {Bogacz}(2010)}]{lebedev2010betatron}%
  \BibitemOpen
  \bibfield  {author} {\bibinfo {author} {\bibfnamefont {V.~A.}\ \bibnamefont
  {Lebedev}}\ and\ \bibinfo {author} {\bibfnamefont {S.}~\bibnamefont
  {Bogacz}},\ }\bibfield  {title} {\bibinfo {title} {Betatron motion with
  coupling of horizontal and vertical degrees of freedom},\ }\href@noop {}
  {\bibfield  {journal} {\bibinfo  {journal} {J. Instrum.}\ }\textbf {\bibinfo
  {volume} {5}}\bibinfo  {number} { (10)},\ \bibinfo {pages}
  {P10010}}\BibitemShut {NoStop}%
\bibitem [{\citenamefont {Lebedev}(2020)}]{lebedev2020ibs}%
  \BibitemOpen
\bibfield  {number} {  }\bibfield  {author} {\bibinfo {author} {\bibfnamefont
  {V.}~\bibnamefont {Lebedev}},\ }\href@noop {} {\emph {\bibinfo {title}
  {Report on Single and Multiple Intrabeam Scattering Measurements in IOTA Ring
  in Fermilab}}},\ \bibinfo {type} {Report No}\ \bibinfo {number}
  {FERMILAB-TM-2750-AD}\ (\bibinfo  {institution} {Fermilab},\ \bibinfo {year}
  {2020})\BibitemShut {NoStop}%
\bibitem [{\citenamefont {Fellenz}\ and\ \citenamefont {Crisp}(1998)}]{WGM}%
  \BibitemOpen
  \bibfield  {author} {\bibinfo {author} {\bibfnamefont {B.}~\bibnamefont
  {Fellenz}}\ and\ \bibinfo {author} {\bibfnamefont {J.}~\bibnamefont
  {Crisp}},\ }\bibfield  {title} {\bibinfo {title} {An improved resistive wall
  monitor},\ }\href {https://doi.org/10.1063/1.57030} {\bibfield  {journal}
  {\bibinfo  {journal} {AIP Conference Proceedings}\ }\textbf {\bibinfo
  {volume} {451}},\ \bibinfo {pages} {446} (\bibinfo {year}
  {1998})}\BibitemShut {NoStop}%
\bibitem [{\citenamefont {Bjorken}\ and\ \citenamefont
  {Mtingwa}(1982)}]{bjorken1982intrabeam}%
  \BibitemOpen
  \bibfield  {author} {\bibinfo {author} {\bibfnamefont {J.~D.}\ \bibnamefont
  {Bjorken}}\ and\ \bibinfo {author} {\bibfnamefont {S.~K.}\ \bibnamefont
  {Mtingwa}},\ }\bibfield  {title} {\bibinfo {title} {Intrabeam scattering},\
  }\href@noop {} {\bibfield  {journal} {\bibinfo  {journal} {Part. Accel.}\
  }\textbf {\bibinfo {volume} {13}},\ \bibinfo {pages} {115} (\bibinfo {year}
  {1982})}\BibitemShut {NoStop}%
\bibitem [{\citenamefont {Nagaitsev}(2005)}]{IBS_Nagaitsev}%
  \BibitemOpen
  \bibfield  {author} {\bibinfo {author} {\bibfnamefont {S.}~\bibnamefont
  {Nagaitsev}},\ }\bibfield  {title} {\bibinfo {title} {Intrabeam scattering
  formulas for fast numerical evaluation},\ }\href
  {https://doi.org/10.1103/PhysRevSTAB.8.064403} {\bibfield  {journal}
  {\bibinfo  {journal} {Phys. Rev. ST Accel. Beams}\ }\textbf {\bibinfo
  {volume} {8}},\ \bibinfo {pages} {064403} (\bibinfo {year}
  {2005})}\BibitemShut {NoStop}%
\bibitem [{\citenamefont {Bernardini}\ \emph {et~al.}(1963)\citenamefont
  {Bernardini}, \citenamefont {Corazza}, \citenamefont {Di~Giugno},
  \citenamefont {Ghigo}, \citenamefont {Haissinski}, \citenamefont {Marin},
  \citenamefont {Querzoli},\ and\ \citenamefont {Touschek}}]{touschek}%
  \BibitemOpen
  \bibfield  {author} {\bibinfo {author} {\bibfnamefont {C.}~\bibnamefont
  {Bernardini}}, \bibinfo {author} {\bibfnamefont {G.~F.}\ \bibnamefont
  {Corazza}}, \bibinfo {author} {\bibfnamefont {G.}~\bibnamefont {Di~Giugno}},
  \bibinfo {author} {\bibfnamefont {G.}~\bibnamefont {Ghigo}}, \bibinfo
  {author} {\bibfnamefont {J.}~\bibnamefont {Haissinski}}, \bibinfo {author}
  {\bibfnamefont {P.}~\bibnamefont {Marin}}, \bibinfo {author} {\bibfnamefont
  {R.}~\bibnamefont {Querzoli}},\ and\ \bibinfo {author} {\bibfnamefont
  {B.}~\bibnamefont {Touschek}},\ }\bibfield  {title} {\bibinfo {title}
  {Lifetime and beam size in a storage ring},\ }\href
  {https://doi.org/10.1103/PhysRevLett.10.407} {\bibfield  {journal} {\bibinfo
  {journal} {Phys. Rev. Lett.}\ }\textbf {\bibinfo {volume} {10}},\ \bibinfo
  {pages} {407} (\bibinfo {year} {1963})}\BibitemShut {NoStop}%
\bibitem [{\citenamefont {Carmignani}(2014)}]{Carmignani:2014qyb}%
  \BibitemOpen
  \bibfield  {author} {\bibinfo {author} {\bibfnamefont {N.}~\bibnamefont
  {Carmignani}},\ }\emph {\bibinfo {title} {{Touschek Lifetime Studies and
  Optimization of the European Synchrotron Radiation Facility}}},\ \href
  {https://doi.org/10.1007/978-3-319-25798-3} {Ph.D. thesis},\ \bibinfo
  {school} {Pisa U.} (\bibinfo {year} {2014}),\ \bibinfo {note}
  {pp.~25--26}\BibitemShut {NoStop}%
\bibitem [{\citenamefont {Lebedev}(2013)}]{Intro_touschek}%
  \BibitemOpen
  \bibfield  {author} {\bibinfo {author} {\bibfnamefont {V.}~\bibnamefont
  {Lebedev}},\ }\bibfield  {title} {\bibinfo {title} {Intrabeam scattering},\
  }in\ \href@noop {} {\emph {\bibinfo {booktitle} {Handbook of Accelerator
  Physics and Engineering}}},\ \bibinfo {editor} {edited by\ \bibinfo {editor}
  {\bibfnamefont {A.}~\bibnamefont {Chao}}, \bibinfo {editor} {\bibfnamefont
  {K.}~\bibnamefont {Mess}}, \bibinfo {editor} {\bibfnamefont {M.}~\bibnamefont
  {Tigner}},\ and\ \bibinfo {editor} {\bibfnamefont {F.}~\bibnamefont
  {Zimmermann}}}\ (\bibinfo  {publisher} {World Scientific},\ \bibinfo {year}
  {2013})\ pp.\ \bibinfo {pages} {155--158}\BibitemShut {NoStop}%
\bibitem [{\citenamefont {Piwinski}(1999)}]{piwinski1999touschek}%
  \BibitemOpen
  \bibfield  {author} {\bibinfo {author} {\bibfnamefont {A.}~\bibnamefont
  {Piwinski}},\ }\href@noop {} {\bibinfo {title} {The {T}ouschek effect in
  strong focusing storage rings}} (\bibinfo {year} {1999}),\ \Eprint
  {https://arxiv.org/abs/physics/9903034} {arXiv:physics/9903034
  [physics.acc-ph]} \BibitemShut {NoStop}%
\bibitem [{\citenamefont {Ha\"{i}ssinski}(1973)}]{haissinski1973exact}%
  \BibitemOpen
  \bibfield  {author} {\bibinfo {author} {\bibfnamefont {J.}~\bibnamefont
  {Ha\"{i}ssinski}},\ }\bibfield  {title} {\bibinfo {title} {Exact longitudinal
  equilibrium distribution of stored electrons in the presence of
  self-fields},\ }\href@noop {} {\bibfield  {journal} {\bibinfo  {journal} {Il
  Nuovo Cimento B (1971-1996)}\ }\textbf {\bibinfo {volume} {18}},\ \bibinfo
  {pages} {72} (\bibinfo {year} {1973})}\BibitemShut {NoStop}%
\bibitem [{\citenamefont {Tenenbaum}\ and\ \citenamefont
  {Shintake}(1999)}]{tenenbaum1999measurement}%
  \BibitemOpen
  \bibfield  {author} {\bibinfo {author} {\bibfnamefont {P.}~\bibnamefont
  {Tenenbaum}}\ and\ \bibinfo {author} {\bibfnamefont {T.}~\bibnamefont
  {Shintake}},\ }\bibfield  {title} {\bibinfo {title} {Measurement of small
  electron-beam spots},\ }\href@noop {} {\bibfield  {journal} {\bibinfo
  {journal} {Annual Review of Nuclear and Particle Science}\ }\textbf {\bibinfo
  {volume} {49}},\ \bibinfo {pages} {125} (\bibinfo {year} {1999})}\BibitemShut
  {NoStop}%
\bibitem [{\citenamefont {Sakai}\ \emph {et~al.}(2001)\citenamefont {Sakai},
  \citenamefont {Honda}, \citenamefont {Sasao}, \citenamefont {Araki},
  \citenamefont {Higashi}, \citenamefont {Okugi}, \citenamefont {Taniguchi},
  \citenamefont {Urakawa},\ and\ \citenamefont
  {Takano}}]{sakai2001measurement}%
  \BibitemOpen
  \bibfield  {author} {\bibinfo {author} {\bibfnamefont {H.}~\bibnamefont
  {Sakai}}, \bibinfo {author} {\bibfnamefont {Y.}~\bibnamefont {Honda}},
  \bibinfo {author} {\bibfnamefont {N.}~\bibnamefont {Sasao}}, \bibinfo
  {author} {\bibfnamefont {S.}~\bibnamefont {Araki}}, \bibinfo {author}
  {\bibfnamefont {Y.}~\bibnamefont {Higashi}}, \bibinfo {author} {\bibfnamefont
  {T.}~\bibnamefont {Okugi}}, \bibinfo {author} {\bibfnamefont
  {T.}~\bibnamefont {Taniguchi}}, \bibinfo {author} {\bibfnamefont
  {J.}~\bibnamefont {Urakawa}},\ and\ \bibinfo {author} {\bibfnamefont
  {M.}~\bibnamefont {Takano}},\ }\bibfield  {title} {\bibinfo {title}
  {Measurement of an electron beam size with a laser wire beam profile
  monitor},\ }\href {https://doi.org/10.1103/PhysRevSTAB.4.022801} {\bibfield
  {journal} {\bibinfo  {journal} {Phys. Rev. ST Accel. Beams}\ }\textbf
  {\bibinfo {volume} {4}},\ \bibinfo {pages} {022801} (\bibinfo {year}
  {2001})}\BibitemShut {NoStop}%
\bibitem [{aps(2019)}]{apsu}%
  \BibitemOpen
  \href {https://www.aps.anl.gov/APS-Upgrade/Documents} {\emph {\bibinfo
  {title} {{Advanced Photon Source Upgrade Project, Final Design Report}}}},\
  \bibinfo {type} {Tech. Rep.}\ \bibinfo {number} {APSU-2.01-RPT-003}\
  (\bibinfo  {institution} {Argonne National Laboratory},\ \bibinfo {address}
  {Lemont, IL},\ \bibinfo {year} {2019})\ \bibinfo {note} {{Chapter~2,
  Table~2.1; Chapter~4, Table~4.62}}\BibitemShut {NoStop}%
\end{thebibliography}%

\end{document}